\documentclass[sigconf]{acmart}
\AtBeginDocument{%
  }

\setcopyright{acmlicensed}
\copyrightyear{2018}
\acmYear{2018}
\acmDOI{XXXXXXX.XXXXXXX}
\acmConference[Conference acronym 'XX]{Make sure to enter the correct
  conference title from your rights confirmation email}{June 03--05,
  2018}{Woodstock, NY}
\acmISBN{978-1-4503-XXXX-X/2018/06}

\usepackage[dvipsnames]{xcolor}
\definecolor{Green}{HTML}{7fc97f}
\definecolor{Purple}{HTML}{beaed4}
\definecolor{Yellow}{HTML}{fdc086}
\usepackage{colortbl} 
\usepackage[table]{xcolor}

\begin{document}

\title{"Nothing about us without us": Perspectives of Global Deaf and Hard-of-hearing Community Members \\ on Sign Language Technologies}

\author{Katherine Atwell}
\authornote{Both authors contributed equally to this research.}
\affiliation{%
  \institution{Northeastern University}
  \country{United States}
}

\author{Saki Imai}
\authornotemark[1]
\affiliation{%
  \institution{Northeastern University}
  \country{United States}
}

\author{Danielle Bragg}
\affiliation{%
  \institution{Microsoft Reserach}
  \country{United States}
}

\author{Malihe Alikhani}
\affiliation{%
  \institution{Northeastern University}
  \country{United States}
}

\renewcommand{\shortauthors}{Atwell et al.}

\begin{abstract}
There is accelerating interest in sign language technologies (SLTs), with increasing attention from both industry and academia. However, the perspectives of Deaf and Hard-of-hearing (DHH) individuals remain marginalized in their development, particularly those outside of the West and in the global South. This paper presents findings from a global, multilingual survey capturing community views on SLTs across a wide range of countries, sign languages, and cultural contexts. While participants recognized the potential of SLTs to support access and independence, many expressed concerns about cultural erasure, inaccurate translation, and hearing-dominated research pipelines. Perceptions of SLTs were shaped by factors including sign language proficiency, policy exposure, and deaf identity. Across regions, participants emphasized the importance of DHH-led design, citing the risk of harm when DHH communities are excluded from technological decision-making. This study offers a novel cross-continental, community-informed analysis of SLTs and concludes with actionable recommendations for researchers, technologists, and policymakers.
\end{abstract}


\begin{teaserfigure}
  \includegraphics[width=\textwidth]{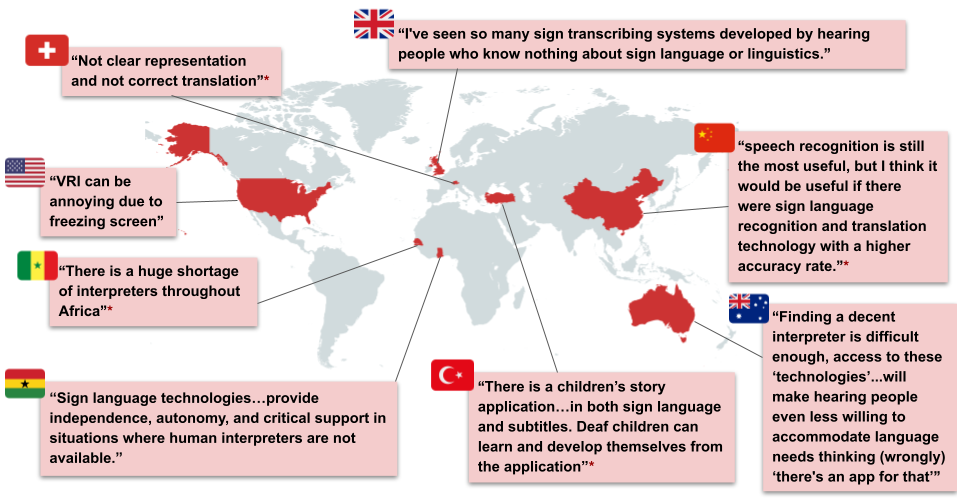}
  \Description{A world map shaded in gray, except for a few countries associated with selected quotes from participants (shaded in red). 8 quotes in total are provided in light red boxes, and for each quote there is a line drawn to a country the quoted participant is from, as well as the country's flag next to the quote. The quotes sampled are: ``VRI can be annoying due to freezing screen" (United States), ``Not clear representation and not correct translation" (Switzerland), ``I've seen so many sign transcribing systems developed by hearing people who know nothing about sign language or linguistics" (United Kingdom), ``Speech recognition is still most useful, but I think it would be useful if there were sign language recognition and translation technology with a higher accuracy rate (China), ``Finding a decent interpreter is difficult enough, access to these `technologies'...will make hearing people even less willing to accommodate language needs thinking (wrongly) `there's an app for that'" (Australia), ``there is a children's story application...in both sign language and subtitles. Deaf children can learn and develop themselves from the application" (T\ddot{u}rkiye), ``Sign language technologies...provide independence, automony, and critical support in situations where human interpreters are not available" (Ghana), and ``There is a huge shortage of interpreters throughout Africa" (Senegal). The sampled responses from Switzerland, China, T\ddot{u}rkiye, and Senegal were translated to English using Google Translate.}
  \caption{A sample of participant opinions about sign language technologies, with countries respondents are from in red. Quotations followed by a red asterisk were translated into English using Google Translate. For participants from multiple countries, a single country is highlighted on the map.}
  \label{fig:teaser}
\end{teaserfigure}

\received{20 February 2007}
\received[revised]{12 March 2009}
\received[accepted]{5 June 2009}

\maketitle

\section{Introduction}


Worldwide, over 1.5 billion people live with hearing loss, and 430 million live with severe hearing loss\footnote{\url{https://www.who.int/health-topics/hearing-loss}}. Sign language is the preferred means of communication for many deaf and hard-of-hearing (DHH) people. There is no universal sign language; rather, there are hundreds of sign language used worldwide, and a variety of DHH cultures and perspectives. 

Emerging sign language technologies (SLTs) have the potential to greatly impact the lives of DHH people around the world. SLT refers to technologies that facilitate communication in a sign language, including (among other use cases) easing language learning, 
communicating with hearing people, or accessing critical services \cite{kushalnagar2019video, huffman2024we}. 
As AI becomes increasingly powerful, SLTs are becoming increasingly viable. As SLTs gain momentum in accessibility and AI research, concerns are also mounting about systems developed without the participation of DHH communities.

Lack of understanding global DHH communities and including them in technology development is a major barrier to developing effective SLTs. 
Although research in SLTs has grown rapidly over the past few years \cite{desai2024systemic}, many of these technologies are unusable for DHH communities due to low performance \cite{yin2021including, o2024deaf} or efforts led by researchers who are unfamiliar with sign languages, Deaf culture, and the needs of DHH communities \cite{yin2021including, fox2023best, desai2024systemic}. 
Though existing systems generally fall short, there is little prior work focused on understanding DHH community perspectives on SLTs. 
While a few existing works survey DHH people regarding certain aspects of SLTs \cite{tran_2023, 10.1145/1361203.1361206, kushalnagar2019video, glasser2022analyzing} these works typically focus on specific regions or sign languages. In particular, DHH people in Western countries or those who communicate in American Sign Language (ASL) are often the focus of these surveys, excluding DHH perspectives from most of the world.

In this work, we survey DHH people around the world to better understand how their lived experiences influence their perspectives on SLTs. Expanding on prior work that focused primarily on western DHH perspectives, we amplify the perspectives of DHH individuals \textit{globally}. We survey individuals from 5 continents regarding their perspectives on SLTs, focusing on three different axes: \textit{identity}, \textit{policy}, and \textit{power}. 



\begin{enumerate}
    \item \textbf{Power:} to what extent to DHH individuals believe SLTs can benefit or harm their communities, particularly in the presence or absence of DHH contributors? (\S\ref{sec:power})
    \item \textbf{Identity:} how do demographic factors (e.g. geographical location or audiological status) and beliefs about deaf identity shape perceptions of SLTs? (\S\ref{sec:identity})
    \item \textbf{Policy:} what policies or programs have influenced DHH individuals' perspectives on SLTs? (\S\ref{sec:policy})
\end{enumerate}

Through our quantitative and thematic analyses, we find that, while most participants are in favor of developing SLTs overall, community members are most often concerned about hearing people profiting from these technologies and a lack of DHH contributors. We find evidence that geographical region may influence perspectives about SLTs. In particular, availability of interpreters appears to shape participants' opinions, with some participants worried that SLTs will replace interpreters while others discuss the potential for these technologies to be utilized when interpretation is not available. Finally, perceptions of the capabilities of these technologies shaped many participants' perspectives, with some participants citing low translation performance or issues with video remote interpreting systems. 

In addition to DHH individuals, who we actively sought out in our recruiting calls, 9 hearing individuals filled out our survey, 5 of whom are sign language researchers or have deaf colleagues. Rather than removing these results, we choose to retain them in order to illustrate the contrast between hearing perspectives and DHH perspectives, even from those who may have some familiarity with these communities.

With these insights, we challenge assumptions that SLTs are inherently empowering, revealing how support is shaped by cultural inclusion, political recognition, infrastructural access, and entrenched inequalities in sign language education.



\section{Background and Related Work}

\subsection{DHH Community}

Rather than representations of written language, sign languages are distinct languages in themselves, with their own grammar that is distinct from written languages. With these distinct languages also come distinct cultures, and many deaf individuals identify with Deaf culture, which refers to communities of signing deaf people. When referring to Deaf culture, a capital D is typically used, and we adopt this practice in this paper. When we use the word ``deaf" in lowercase, we refer to those who are not able to hear. However, when we use the term ``Deaf" with a capital D, we are referring to Deaf culture. The one exception to this is when reporting results, as we capitalize Deaf, Hard-of-hearing, and Hearing when referring to the different audiological statuses of participants in this work. 

In total, there are over 200 sign languages \footnote{\url{https://wfdeaf.org/our-work/}} alone. Different cultures are also present among signers who use the same language. An example is the Deaf Black community \cite{aramburo1989sociolinguistic}. These cultural differences can manifest linguistically; for instance, Black signers in the United States often use a distinct dialect of sign language, known as Black ASL \cite{mccaskill2011hidden}. 

Unfortunately, despite the large number of distinct sign languages globally, many are still unrecognized as languages by the countries in which they are used. Out of 193 total countries, only 78 provide legal recognition of sign languages\footnote{\url{https://wfdeaf.org/news/the-legal-recognition-of-national-sign-languages/}}, which is a critical need for community members \cite{de2016power}. Many DHH individuals, even those who have been deaf since birth, are not properly educated in sign languages, which can have a lasting impact on language development \cite{mayberry2024impoverished}. Further, health inequities persist between deaf and hearing individuals, which are perpetuated by lack of access \cite{barnett2011deaf}.




\subsection{Sign Language Technology}

Sign language technology (SLT) broadly refers to software and hardware interfaces that support communication in a sign language.  
SLTs have been applied for a number of purposes, including 
education \cite{mokhtar2015learning, sogani2025interaction}, and healthcare settings \cite{kushalnagar2019video}. Some applications of SLTs include apps for learning sign language \cite{mokhtar2015learning, sogani2025interaction}, video remote interpreting systems \cite{kushalnagar2019video}, and personal assistants that understand sign languages \cite{glasser2022analyzing}.

Automatic sign language processing research is a growing field that can enable increasingly powerful SLTs. It is an established area in computer vision (CV) and  natural language processing (NLP), and also studied in human-computer interaction (HCI). It typically focuses on sign language \textit{detection} \cite{8683257}, \textit{identification} \cite{gebre2013automatic}, \textit{segmentation} \cite{renz2021sign}, \textit{recognition} (individual or continuous) \cite{camgoz2020sign}, \textit{translation} \cite{muller2023findings}, or \textit{production} \cite{inan2022modeling, yin2021including}. 
Datasets and resources for documenting, learning, or processing sign languages include sign language dictionaries \cite{spreadthesign}, lexical databases \cite{caselli2017asl}, and corpora that support isolated sign recognition \cite{desai2023asl, kezar2023sem} and continuous sign recognition \cite{yin2024asl, duarte2021how2sign, hanke-etal-2020-extending}. A prior interdisciplinary review of this space includes a call for more Deaf involvement \cite{bragg2019sign}.




Despite active research and development on SLTs, many issues have been identified with current systems. These issues are often related to either 1) low performance for automatic recognition, translation, or production of sign languages \cite{o2024deaf} or 2) misalignment between developed systems and the needs of DHH communities. The first issue is in direct contrast with AI systems for written languages, which have achieved very high performance and are shown to improve human productivity \footnote{\url{https://www.brookings.edu/articles/breaking-the-ai-mirror/}} \footnote{\url{https://www.oecd.org/en/publications/the-impact-of-artificial-intelligence-on-productivity-distribution-and-growth_8d900037-en.html}}. Performance issues often result from a lack of training data \cite{yin2021including}, and can be addressed by the creation of high-quality datasets for sign language processing \cite{desai2023asl, kezar2023sem}. Many sign language models also lack linguistic foundation in sign languages, which are distinct from written languages and structured differently \cite{desai2024systemic}. The second issue, misalignment between SLTs and community needs, occurs when hearing researchers and technologists develop systems without involving community members or understanding community needs. \cite{fox2023best}. 


\subsection{Understanding DHH Perspectives}

To ensure that SLTs are informed by community needs and grounded in linguistic features of sign languages, DHH individuals should be involved in every phase of research and system development and hearing researchers should understand community needs and linguistic properties of sign languages \cite{fox2023best}. To showcase community needs, existing works have surveyed DHH community members regarding sign language translation \cite{tran_2023}, quality of generated signs \cite{10.1145/1361203.1361206}, video remote interpreting systems \cite{kushalnagar2019video}, and personal assistants that understand sign language \cite{glasser2022analyzing}. 
Many of these works survey only from signing individuals who use a particular sign language, which is often American Sign Language (ASL), or are from a particular country or region (often western countries). To supplement these works and center global community perspectives, we survey members of DHH communities around the world, with the goal of amplifying the perspectives of DHH individuals globally.

Many of these inequities are rooted \emph{audism}, which is a persistent issue encountered by DHH people. Audism is defined as ``the notion that one is superior based on one's ability to hear or behave in the manner of one who hears" \cite{humphries1975}. Without actively involving community members in research and design processes, there is a risk of perpetuating audism in sign language technologies. For instance, signing gloves designed by hearing researchers were rejected by many community members, partially due to concerns about audism \cite{uw_signaloud_2016}. 
\section{Methods}
We conducted a global survey to collect diverse perspectives from DHH individuals on deafness, SLTs, and related policies. The study was approved by the Institutional Review Board (IRB) of the participating institution. 
This section describes the recruitment methods, survey structure, and respondents of our study. We also note the limitations of our study and include a positionality statement.
\subsection{Recruitment}

To ensure broad participation across geographic, linguistic, and cultural contexts, we distributed the survey to over 120 individuals and organizations working with DHH communities across 35 countries in regions including North America, South America, Europe, Africa, the Middle East, South Asia, East Asia, and Oceania. 
We also conducted extensive outreach through social media platforms, posting in 39 Facebook groups focused on Deaf communities in various countries, 3 relevant subreddits, and LinkedIn, X, and Bluesky. Additionally, physical flyers were distributed at a leading research university specializing in deaf education.
The recruitment materials and survey were available in a variety of written languages (detailed below), and our recruitment message stated that we would do our best to provide interpretation support upon request.



\subsection{Survey Structure}

The survey was published on GitHub\footnote{The link will be disclosed after the review process to maintain anonymity.} and distributed through a Google Forms link, available in the following written languages: English, Spanish, French, Chinese (Traditional), Chinese (Simplified), Turkish, Farsi, Arabic, Hindi, and Japanese. To translate the survey, we used ChatGPT to generate initial translations, then checked each translation with a native speaker. 

Based on feedback we received after initial release of the survey, we simplified and structured the text of the consent form to make it more accessible to participants who were less comfortable with written languages. 
After consent, the survey consisted of three main sections, combining multiple-choice and open-ended questions:
\begin{enumerate}
    \item \textit{Views of Sign Language Technology} – This section explored participants’ concerns, perceived benefits, and overall attitudes toward the development of sign language-related technologies.
    \item \textit{View of Deafness} - Participants were asked about their views on deafness (e.g., as a cultural identity or a disability) and any local policies or programs they believed have benefited or harmed the Deaf community.
    \item \textit{Demographics} - This section collected general background information to better understand the diversity of participants. Specifically, we asked about age, gender, country of residence, audiological status (e.g., Deaf, Hard of hearing, Hearing), sign language proficiency, and area of residence (e.g., urban, suburban, rural). No personally identifiable information was collected at any stage of the study, including names and email addresses, to ensure that participants' responses could not be traced back to any individual.
\end{enumerate}

To incentivize participants and show appreciation, for each of the first 1,000 completed surveys, a \$5 donation was made to the World Federation of the Deaf (WFD). Please refer to Appendix~\ref{subsec:survey} for the complete list of survey questions.

\subsection{Data Analysis}
For the multiple-choice survey questions, we report descriptive statistics, including proportions of responses across different subsets of participants (for instance, by audiological status, sign language proficiency, and residential context). Where relevant, we draw comparisons across groups to highlight trends and disparities in perspectives.

For the four open-ended questions, we conducted the initial coding and analysis process using Braun and Clarke’s constructivist thematic analysis approach \cite{braun2006using}. Coding was conducted inductively, allowing themes to emerge from the data. We prioritized preserving participants’ opinions by integrating direct quotes to illustrate each theme.

\subsection{Participants}
A total of 42 respondents completed our survey between June 2024 and May 2025. Table~\ref{tab:freq} summarizes participants' audiological status alongside the continents in which they have lived. Rather than limiting participants to a single country, we asked them to select all countries where they had resided. We adopted this approach because we believe that lived experience across different regions can shape individuals’ perspectives on SLTs and deafness. Because participants could list multiple countries, we divided their response proportionally across the corresponding continents, which resulted in fractional values in the table. 

Participant demographics were as follows: gender — female (24), male (16), and other (2); age ranged from 26 to 65 years ($\mu = 43$, $\sigma = 12$). Participants also self-reported their sign language proficiency using the American Sign Language Proficiency Interview (ASLPI) scale \cite{gallaudet_aslpi}. Table~\ref{tab:asl_proficiency} summarizes the distribution of ASL proficiency levels across participants' reported audiological status. 

While our recruitment efforts were primarily directed at DHH individuals, our final sample included nine hearing participants. Of these, four identified as sign language researchers, one reported working closely with a deaf colleague. The remaining hearing participants did not clearly indicate community involvement.

\begin{table*}[h]
  \centering
  \caption{Summary of study respondents by continent and audiological status (fractional counts)}
  \label{tab:freq}
  \begin{tabular}{l|ccccc|c}
    \toprule
     & \textbf{Africa} & \textbf{Asia} & \textbf{Europe} & \textbf{North America} & \textbf{Oceania} & \textbf{Total} \\
    \midrule
    Deaf            
      & \cellcolor[HTML]{99BFF9}3.5
      & \cellcolor[HTML]{E6EEFC}1.0
      & \cellcolor[HTML]{7CAEF7}5.0
      & \cellcolor[HTML]{2F72E9}11.0
      & \cellcolor[HTML]{F3F7FE}0.5
      & 21 \\
    Hard-of-hearing  
      & \cellcolor[HTML]{A7C7FA}2.5
      & \cellcolor[HTML]{D1E0FA}1.4
      & \cellcolor[HTML]{E8F0FD}0.7
      & \cellcolor[HTML]{8EB9F8}4.2
      & \cellcolor[HTML]{6DA4F6}3.2
      & 12 \\
    Hearing          
      & \cellcolor[HTML]{FFFFFF}0
      & \cellcolor[HTML]{5E9BF4}3.83
      & \cellcolor[HTML]{A4C5FA}1.67
      & \cellcolor[HTML]{B6CFFA}2.0
      & \cellcolor[HTML]{BFD4FA}1.5
      & 9 \\
    \bottomrule
  \end{tabular}
\end{table*}

\begin{table}[h!]
    \centering
    \caption{SL Proficiency Levels (0–5) by Audiological Status}
    \begin{tabular}{lcccccc}
    \toprule
     & \textbf{0} & \textbf{1} & \textbf{2} & \textbf{3} & \textbf{4} & \textbf{5} \\
    \midrule
    Deaf           
      & \cellcolor[HTML]{FFFFFF}0
      & \cellcolor[HTML]{FFFFFF}0
      & \cellcolor[HTML]{DDE8FB}3
      & \cellcolor[HTML]{D1E0FA}2
      & \cellcolor[HTML]{C3D7F9}3
      & \cellcolor[HTML]{85ACF4}13 \\
    Hard-of-hearing 
      & \cellcolor[HTML]{FFFFFF}0
      & \cellcolor[HTML]{CFE0FA}3
      & \cellcolor[HTML]{FFFFFF}0
      & \cellcolor[HTML]{C3D7F9}4
      & \cellcolor[HTML]{D1E0FA}3
      & \cellcolor[HTML]{E6EEFC}2 \\
    Hearing        
      & \cellcolor[HTML]{C3D7F9}4
      & \cellcolor[HTML]{D1E0FA}3
      & \cellcolor[HTML]{E6EEFC}1
      & \cellcolor[HTML]{FFFFFF}0
      & \cellcolor[HTML]{E6EEFC}1
      & \cellcolor[HTML]{FFFFFF}0 \\
    \bottomrule
    \end{tabular}
    \label{tab:asl_proficiency}
\end{table}


\subsection{Limitations}
Despite our efforts to reach a diverse population, the survey's distribution channels may have introduced selection bias. Participants with more internet access or social media accounts were more likely to be represented, potentially limiting the generalizability of the findings. Additionally, the survey languages may have excluded speakers of unsupported languages, and the use of written forms instead of signed surveys may have influenced comprehension.
For the responses of the survey that weren't in English, we used Google Translate for analysis pre-processing.

\subsubsection{Positionality Statement}
All authors of this paper are hearing researchers. We recognize that our positionality has inevitably shaped how we approach, analyze and interpret the findings \cite{secules2021positionality}. While none of us identify as DHH, each of us has prior experience working closely with DHH researchers and community members in various capacities. This includes studying signed languages, attending Deaf community events, and many previous works published with DHH collaborators. These relationships have informed our understanding and approach, and we have also consulted existing literature and position papers written by DHH researchers to ensure our framing aligns with community priorities. These steps were taken to remain accountable to DHH perspectives and reduce the risk of perpetuating hearing-centered assumptions.

\section{Results}
\subsection{Concerns for Sign Language Technologies (SLTs)}
\label{sec:power}
In this section, we present participants’ concerns regarding SLTs.
\subsubsection{Concerns}
\label{sec:concerns}
\begin{figure}[h]
  \centering
  \includegraphics[width=\linewidth]{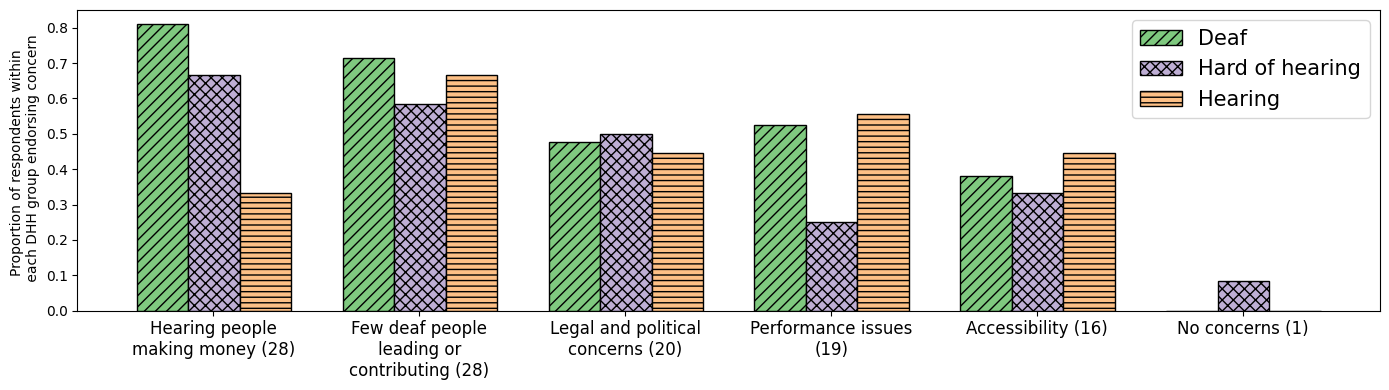}
  \caption{Concerns about SLTs by audiological group. The x-axis lists specific concerns endorsed by participants, with number of mentions in parentheses. Bars show the proportion of respondents within each group who selected each concerns.}
  \Description{Bar chart showing concerns about SLTs among Deaf, Hard-of-hearing, Hearing participants. Each concern is listed on the x-axis with its total number of mentions in parentheses. For example, 81\% of Deaf participants expressed concern about hearing people making money from SLTs, compared to 67\% of Hard of hearing and 33\% of Hearing participants. The figure uses uses three patterns and colors to distinguish groups: green diagonal lines for Deaf, purple crosshatch for Hard of hearing, and orange horizontal lines for Hearing.}
  \label{fig:concerns}
\end{figure}

We begin with the multiple choice survey findings, which allowed participants to select all concerns that applied from a predefined list (with an additional free-response option). These multiple-choice items were developed based on prior work identifying key concerns among Deaf communities in the United States \cite{tran_2023}.
Figure~\ref{fig:concerns} presents both the total number of respondents endorsing each concern (in the x-axis), as well as the proportion of respondents within each participant group (\colorbox{Green}{Deaf}, \colorbox{Purple}{Hard-of-hearing}, \colorbox{Yellow}{Hearing}) to help address unequal group sizes.

Overall, the top two concerns selected were \emph{“Hearing people making money”} (28 participants) and \emph{“Few deaf people leading or contributing”} (28 participants). These concerns were especially prominent among deaf participants, with 81\% endorsing the first and 71\% endorsing the second. In contrast, only 33\% of hearing participants selected the concern about hearing people profiting. 

DHH respondents generally shared similar concerns, with one key difference: concerns about \emph{"System performance issues"} were selected by over half of Deaf participants (52\%) but only a quarter of Hard-of-hearing participants (25\%). Some hard-of-hearing respondents mentioned in our survey that captioning systems met many of their accessibility needs, which may explain some of this discrepancy. Hard-of-hearing individuals are more likely to use spoken/written language, which can be supported with captioning, whereas Deaf individuals are more likely to prefer sign language relative to hard-of-hearing individuals.  Written languages are typically better resourced than sign languages, and speech recognition is more accurate than sign language translation, which is currently not viable for real-world use. Past research has also shown that higher degrees of hearing loss may be associated with greater difficulties communicating in written languages due to language deprivation in early life \cite{10.1093/deafed/eni026}. 

Other concerns such as \emph{legal/political implications} and \emph{accessibility} were selected by a moderate number of participants across all groups, with endorsement rates ranging from one-third to half of respondents within each subgroup. Only one respondent selected \emph{“No concerns”}, indicating pervasive concerns about SLTs across the sample. Deaf participants, in particular, consistently endorsed a greater number of concerns than Hard-of-hearing and Hearing participants. This suggests that Deaf individuals may have a more critical or cautious stance toward SLTs. 

In addition to analyzing responses by audiological status, we conducted an analysis based on participants' self-reported sign language proficiency. As shown in Table~\ref{tab:asl_proficiency}, participants rated their sign language proficiency on a scale from 0 to 5. For analysis, we grouped responses into three proficiency levels: low (0-2), moderate (3-4), and high (5).  

\begin{figure}[h]
  \centering
  \includegraphics[width=\linewidth]{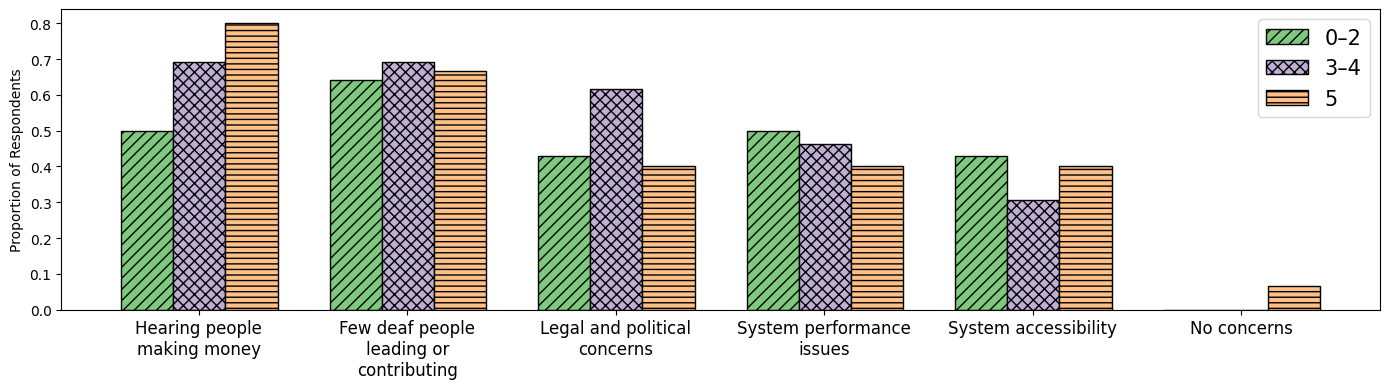}
  \caption{Concerns about SLTs grouped by participants' sign language proficiency. Participants were grouped into low (0-2), moderate (3-4), and high (5) proficiency categories. Bars show the proportion of respondents within each group who endorsed each concern.}
  \Description{Bar chart showing the proportion of respondents endorsing key concerns about SLTs, grouped by self-rated sign language proficiency: 0-2 (green diagonal lines), 3-4 (purple crosshatch), 5 (orange horizontal lines). The most frequently cited concerns among high proficiency participants were "Hearing people making money" (80\%) and "Few deaf people leading or contributing" (over 65\%). Lower proficiency participants expressed fewer concerns overall.}
  \label{fig:concerns_asl}
\end{figure}

Figure~\ref{fig:concerns_asl} presents the proportion of participants within each proficiency group who endorsed key concerns. The pattern of concerns by sign language proficiency closely mirrors the trend observed across audiological status groups. For instance, participants with higher sign language proficiency expressed a greater number of concerns, similar to the trend seen among deaf participants. 


These findings suggest that perspectives on SLTs may be shaped not only by one's audiological status but also by the depth and nature of their engagement with signed communication. Participants with higher sign language proficiency may be more attuned to the cultural and communicative implications of signing that lead to heightened awareness of potential harms, misrepresentation, or exclusion. While sign language proficiency and audiological status are correlated in this sample, the consistency of concern patterns across both groups suggests lived language experience as an important lens to evaluate the risks of SLTs.

\subsubsection{Thematic analysis about concerns}
To contextualize the multiple-choice results above, we conducted thematic analysis of particpants' open-ended responses on concerns about SLT. 

\begin{table*}[h]
  \centering
  \renewcommand{\arraystretch}{1.3}
  \caption{Themes and representative quotes from responses to: “Briefly describe the life experiences that have influenced your perspective on sign language technologies.” The quotes followed by an asterisk were translated into English using Google Translate.}
  \begin{tabular}{p{4cm} p{9cm}}
    \toprule
    \textbf{Theme} & \textbf{Representative Quote} \\
    \midrule
    Skepticism \& accuracy concerns (12) & "I also see nothing comparable that gives me hope that any technology could accurately capture and comprehend the complexity of a manual language." (P24) \\
    Hesitation due to hearing-dominated research (3) & "I've seen so many sign transcribing systems developed by hearing people who know nothing about sign language or linguistics." (P26) \\
    \bottomrule
  \end{tabular}
  \label{tab:lived-experiences-concerns}
\end{table*}

The most cited concern was \emph{skepticism and concerns over accuracy} of current SLTs (n=12). Respondents highlighted the limitations of current systems in capturing the complexity and speed of natural sign language (P9, P16, P20, P24, P31). These concerns were often tied to frustrations with video remote interpreting services that suffered from technical issues like slow internet connections (P33, P34, P37, P38), as well as the limitations of spoken language transcription systems that frequently produce inaccurate captions (P6, P25). While some participants focused their critiques on the current limitations of these technologies:
\begin{quote}
    \textit{``No experience with sign language tech. But seeing how inaccurate automatic captions are, I don’t trust any sign language tech.''} (P25)
\end{quote}
Others expressed more fundamental reservations about the use of non-human solutions for interpretation at all:
\begin{quote}
    \textit{``I don't want to rely on non human technology for things as important and complex as interpretation. This screams 'making it easier and cheaper for hearing people without regard for deaf people'''} (P16)
\end{quote}

Some participants (n=3) highlighted the need for deaf-led research and development, expressing \emph{hesitation about projects dominated by hearing researchers} with limited understanding of the community’s needs. These participants questioned whether current projects truly reflect the needs and lived experiences of DHH users. One participant, a deaf researcher, reflected on their own experience of exclusion:
\begin{quote}
    \textit{``As a deaf researcher I loved doing work in this field but I feel like I slowly got pushed out and I can’t name a deaf researcher who is doing this...the deaf perspective, even if you grew up oral, is very different from a hearing one.''} (P20)
\end{quote}
Others raised concerns about the broader consequences of hearing-led development, suggesting that poorly designed technologies risk further marginalizing deaf users:
\begin{quote}
    \textit{``Access to these ‘technologies’—which are generally not useful and created as vanity projects by hearing dilettantes with no understanding of the deaf community—will make hearing people even less willing to accommodate language needs, thinking (wrongly) ‘there’s an app for that.’''} (P23)\\
    \textit{``I've seen so many sign transcribing systems developed by hearing people who know nothing about sign language or linguistics. It's so wrong, 'cos they are funded millions yet have no outcomes that I am aware of that benefit real Deaf people who use sign language.''} (P26)
\end{quote}





\subsection{Perceived Benefits for SLTs}
In this section, we present participants’ perceived benefits regarding SLTs.
\subsubsection{Perceived Benefits}
\begin{figure}[h]
  \centering
  \includegraphics[width=\linewidth]{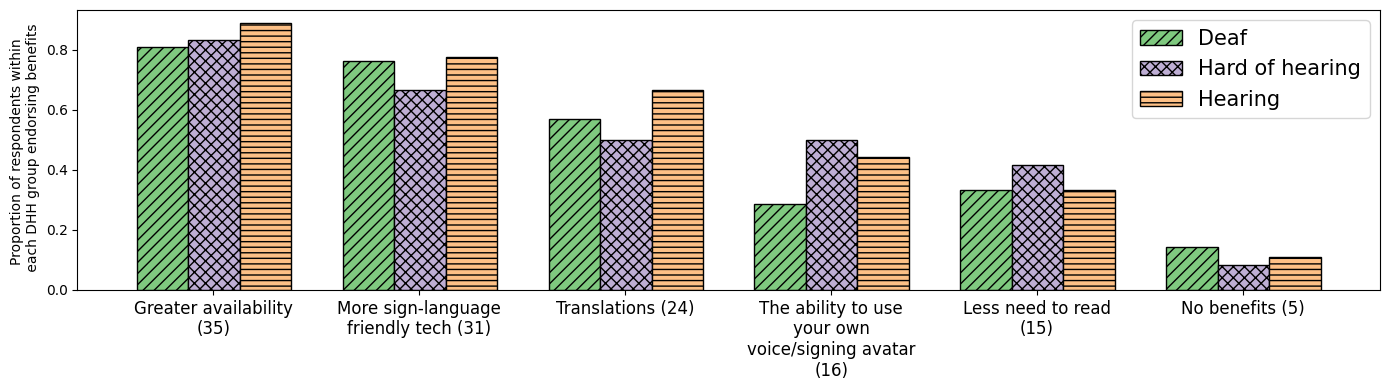}
  \caption{Perceived benefits of SLTs by audiological group. Bars represent the proportion of respondents within each group (Deaf, Hard-of-hearing, Hearing) who endorsed each potential benefit. }
  \Description{Bar chart displaying proportions of Deaf, Hard-of-hearing, and Hearing participants who endorsed each predefined benefit of SLTs. Six benefit categories are shown on the x-axis, each labeled with the total number of mentions in parentheses. The most commonly endorsed benefits across all groups were “Greater availability” (over 75\% in each group) and “More sign-language friendly tech.” Hearing participants showed the highest endorsement for “Translations” (67\%). }
  \label{fig:benefits}
\end{figure}

Participants were also asked to identify the potential benefits they associate with SLTs by selecting from a multiple-choice list (with a write-in option). Figure~\ref{fig:benefits} presents the proportion of participants within each audiological group who endorsed various benefits. Overall, responses reflect the positive impact SLTs could have in terms of improving accessibility and communication.

Across all groups, the two most widely endorsed benefits were \emph{“Greater availability of services”} and \emph{“More sign-language friendly technologies”}. These items were selected by over 80\% of Deaf and Hard-of-hearing participants and by an even higher proportion (89\%) of Hearing participants for the former. This agreement suggests that participants view SLTs as a tool for expanding access to essential services and enhancing communication infrastructure.

Another highly endorsed benefit, especially among hearing participants, was \emph{“Translations between primary sign language and spoken/written language”}, selected by 67\% of hearing participants, substantially higher than among Deaf (57\%) and Hard-of-hearing respondents (50\%). This may reflect the hearing group's interest in bridging communication gaps, while DHH participants may prioritize other aspects of inclusion, such as cultural or community-based usage. This result also reflects the overemphasis on sign language translation research by hearing community to “build a bridge” for the deaf person to cross \cite{fox2023best, desai2024systemic, Forshay2016}.

Perceived benefits related to personalization and usability, such as \emph{“Less need to read”} and \emph{“The ability to use your own voice/signing avatar”} were endorsed at more moderate rates. These items were selected by 30–50\% of respondents across all groups, suggesting that while these benefits are appreciated, they may not be viewed as essential across the community. 

A small proportion of participants in each group selected \emph{“No benefits”}, with 14\% of deaf, 8\% of hard-of-hearing, and 11\% of hearing participants indicating skepticism about the value of SLTs. These responses may reflect a concern that such technologies could be underdeveloped, poorly implemented, or fail to meaningfully include deaf perspectives, a theme echoed in the concerns reported in Section~\ref{sec:concerns}.

These results show that while participants generally see SLTs as promising tools for enhancing accessibility, the specific benefits prioritized vary slightly across audiological groups.

\begin{table*}[h]
  \centering
  \renewcommand{\arraystretch}{1.3}
  \caption{Themes and representative quotes from responses to: “Briefly describe the life experiences that have influenced your perspective on sign language technologies.” The quotes followed by an asterisk were translated into English using Google Translate.}
  \begin{tabular}{p{4cm} p{9cm}}
    \toprule
    \textbf{Theme} & \textbf{Representative Quote} \\
    \midrule
    Positive potentials of SLTs (12) & "More doctor’s offices provide VRI which saves time compared to requesting an in-person interpreter in advance as that process can push back the appointment." (P33) \\
    Empowerment \& inclusion of DHH individuals (4) & "I strongly believe that advancing these technologies will empower individuals, foster inclusivity, and enhance quality of life." (P29) \\
    \bottomrule
  \end{tabular}
  \label{tab:lived-experiences-benefits}
\end{table*}

\subsubsection{Thematic analysis about benefits}
To contextualize the multiple-choice results, we conducted thematic analysis of participants’ open-ended reflections on SLT benefits. These themes highlight how SLTs are valued not only for accessibility but also for empowerment, visibility, and cultural affirmation, which align with the results presented in \cite{Vogel}.

The most commonly expressed themes included the \emph{positive potential of SLTs} (n=12), particularly in improving communication and supporting accessibility in everyday and high-stakes contexts. These responses illustrate the diverse domains in which SLTs are already making an impact, from personal interactions (P7) to education, emergency response (P10), and healthcare (P33).
Several participants pointed to the utility of SLTs in time-sensitive or high-pressure scenarios. 
While not strictly a form of SLT, automatic speech recognition (ASR) was also cited as a transformative tool for increasing accessibility. As one participant noted:
\begin{quote} 
    \textit{“ASR has changed my life. YouTube used to be a site I ignored and then became accessible with constantly improved captioning... We now have pervasive (imperfect) accessibility anywhere anytime. I think machine signing will be similar.”} (P21) 
\end{quote}

A few participants (n=4) reflected on how SLTs could contribute to broader \emph{empowerment and inclusion of DHH individuals}, especially when designed to increase access, autonomy, and societal visibility. These respondents emphasized that SLTs, when accurate and culturally sensitive, hold promise not just for accessibility, but for societal change. For instance, several participants highlighted the role of SLTs in promoting independence and reducing reliance on human interpreters:
\begin{quote}
    \textit{``They provide independence, autonomy, and critical support in situations where human interpreters are not available.''} (P7)
\end{quote}
Others envisioned their potential for future generations:
\begin{quote}
    \textit{``I strongly believe that advancing these technologies will empower individuals, foster inclusivity, and enhance the quality of life for many.''} (P29)
\end{quote}
Another participant emphasized the symbolic power of SLTs in elevating the status and visibility of sign language and deaf culture:
\begin{quote}
    \textit{``SL tech will contribute to a more inclusive world, not just by enhancing accessibility but also by increasing visibility of deaf people and of sign language as a natural human language.''} (P18)
\end{quote}
P29 further emphasized that for such technologies to succeed, they must be developed with care: \emph{“Ensuring that these tools are accurate, culturally sensitive, and widely accessible is crucial for their success.”}
\subsection{Power}
\label{sec:power}
We explore participants’ perceptions of power, specifically, whether SLTs are seen as enabling or limiting educational access.
These included life experiences that informed perspectives on SLTs and whether such technologies \textit{should} be developed.

\subsubsection{Life Experiences Informing Perspectives on SLTs}
This section draws from responses to an open-ended question aimed at understanding personal experiences that shaped participants’ attitudes toward SLTs. Participants were asked to describe the life experiences that have influenced their perspective on sign language technologies. Participants expressed a strong desire for SLTs in contexts where communication barriers remain, while cautioning against the idea that SLTs could replace human interpreters.   

\emph{Interpreter shortages and limited access} (n=9) were frequently cited, but participants’ reflections also pointed to deeper structural issues shaping communication equity. Participants noted both the unavailability of interpreters (P6, P9, P12) and instances where interpreters imposed financial burdens by charging high fees (P35). These challenges were not framed merely as logistical inconveniences, but as systematic exclusions that impacted access to education (P8) and civic participation (P29). Some participants saw SLTs as having the potential to mitigate these inequities. For instance, one participant reflected:

\begin{quote}
    \textit{``There have been instances where the lack of an interpreter or real-time transcription made it difficult to fully engage in discussions...
    sign language technologies could have bridged the gap, ensuring equal access to communication and participation.''} (P29)
\end{quote}
Another participant echoed these concerns, by describing how interpreter inaccessibility combined with high costs affected their own experiences:
\begin{quote}
    \textit{``Interpreters exploit our need for them by imposing fees for sign language interpretation services, which may be exorbitant in value, in addition to their lack of permanent availability, which results in not providing all services to me or my hearing children when they were children.''} (P35)
\end{quote}

A small number of participants (n=2) emphasized that SLTs should be designed to \emph{augment, not replace}, human interpreters. These responses reflected a belief that while SLTs can enhance accessibility, they should not be viewed as substitutes for human expertise (P17) or community-driven language learning (P30). 

\subsubsection{Should SLTs be developed?}
\label{sec:support}
Table~\ref{tab:SLtech_dev} presents the distribution of responses to a multiple-choice question: ``Thinking about your answers about the possible benefits and harms of sign language technologies, should sign language technologies be developed?'' Participants could select from binary options (\emph{Yes}/\emph{No}) and an additional "Other" option for open-ended responses.


\begin{table}[h]
  \centering
  \caption{Participants' responses to ``Thinking about your answers about the possible benefits and harms of sign language technologies, should sign language technologies be developed?''}
  \label{tab:SLtech_dev}
  \begin{tabular}{lccc}
    \toprule
    & \textbf{Yes} & \textbf{No} & \textbf{Other} \\
    \midrule
    \textbf{Deaf} 
      & \cellcolor[HTML]{3E7EEB}61.9\%
      & \cellcolor[HTML]{F0F5FE}4.8\%
      & \cellcolor[HTML]{7CAEF7}33.3\% \\
    \textbf{Hard-of-hearing} 
      & \cellcolor[HTML]{2F72E9}66.7\%
      & \cellcolor[HTML]{FFFFFF}0.0\%
      & \cellcolor[HTML]{7CAEF7}33.3\% \\
    \textbf{Hearing} 
      & \cellcolor[HTML]{1143A8}\textcolor{white}{88.9\%}
      & \cellcolor[HTML]{D9E6FB}11.1\%
      & \cellcolor[HTML]{FFFFFF}0.0\% \\
    \bottomrule
  \end{tabular}
\end{table}

While a majority across all groups supported the development of SLTs (61.9\% of deaf, 66.7\% of hard-of-hearing, 88.9\% of hearing participants), DHH respondents were more likely to express conditional support or reservations by selecting the "Other" option (33.3\%). In contrast, none of the hearing respondents chose this alternative.

The open-ended responses offered by these participants revealed complex ethical and societal implications of SLTs. As shown in Table~\ref{tab:participant-responses}, participants' perspectives reflect three major themes. Some expressed \emph{conditional approval}, emphasizing the importance of ethical oversight and deaf-led development:
\begin{quote}
    \textit{``It depends. This isn't a black and white issue. `Nothing about us without us.' Many technologies don't involve deaf people and actually make it worse for us. So that kind of tech, no. Others that are deaf led, probably ''} (P16)
\end{quote}
Others emphasized the \emph{inevitability} of the development of such technologies:
\begin{quote}
    \textit{``It's not a matter of if it will be developed, it WILL be developed, just a matter of who will lead the charge.''} (P13)\\
\end{quote}
Additionally, some expressed \emph{concerns about the cultural consequences} of relying too heavily on such technologies:
\begin{quote}
    \textit{``Accuracy can be lost due to AI or misinformation, and strong community dissolves due to isolation of relying on technology. Therefore, no strong Deaf developments come from technology.''} (P31)\\
\end{quote}
While many see their potential benefits, there is a clear call for SLT development to be community-led, transparent, and guided by strong ethical frameworks. The perspective that ``\textit{nothing about us without us}'' (P16) was a recurring theme, and this reinforces the need for deaf leadership in shaping these technologies.

\begin{table*}[h]
  \centering
  \renewcommand{\arraystretch}{1.2}
  \caption{Representative quotes reflecting different perspectives on the development of sign language technologies.}
  \begin{tabular}{p{3.5cm} p{9cm}}
    \toprule
    \textbf{Theme} & \textbf{Representative Quote} \\
    \midrule
    Conditional support & ``I am still back and forth on this, but if Deaf researchers are involved at each step, I will say yes.'' (P20) \\
    Inevitability & ``It's not a matter of if it will be developed, it WILL be developed, just a matter of who will lead the charge.'' (P13) \\
    Cultural concerns & ``This does not replace interpreters, and youth should still be given the option to learn sign languages and not be forced to use devices.'' (P30) \\
    \bottomrule
  \end{tabular}
  \label{tab:participant-responses}
\end{table*}

\subsection{Impact of Identity on SLT Perspective}
\label{sec:identity}
\subsubsection{Opinion of SLT}
Participants were asked: “Has your life experience shifted your opinion of sign language technologies in a positive, negative, or neutral way overall?” As shown in Table~\ref{tab:lifeexp_shift}, responses were mixed across groups. 

Those who identified deafness primarily as a \textit{cultural identity} expressed the least enthusiasm for SLTs: only 30.8\% reported a positive shift in perspective, compared to more than half who described their stance as neutral (53.8\%) and 15.4\% as negative. In contrast, participants who conceptualized deafness as a \textit{disability}, or as \textit{both a disability and a cultural identity}, were more favorable toward technological development.

These results suggest that cultural framings of deafness may foster skepticism toward SLTs, while disability-oriented framings may increase openness to technological intervention. This result aligns with the prior research that participants who view deafness as a cultural identity may regard SLTs as unnecessary or potentially undermining cultural preservation, whereas those who emphasize disability may see them as tools for accessibility and inclusion \cite{jimenez2021psychosocial}.

\begin{table}[h]
  \centering
  \caption{Responses to ``Has your life experience shifted your opinion of sign language technologies in a positive, negative, or neutral way overall?''}
  \label{tab:lifeexp_shift}
  \begin{tabular}{lccc}
    \toprule
    & \textbf{Positive} & \textbf{Neutral} & \textbf{Negative} \\
    \midrule
    \textbf{A cultural identity} 
      & \cellcolor[HTML]{99BFF9}30.8\%
      & \cellcolor[HTML]{2F72E9}53.8\%
      & \cellcolor[HTML]{D9E6FB}15.4\% \\
    \textbf{A disability} 
      & \cellcolor[HTML]{3E7EEB}50.0\%
      & \cellcolor[HTML]{FFFFFF}0.0\%
      & \cellcolor[HTML]{3E7EEB}50.0\% \\
    \textbf{Both of the above} 
      & \cellcolor[HTML]{4C8AEF}46.2\%
      & \cellcolor[HTML]{75ABF6}34.6\%
      & \cellcolor[HTML]{CEE0FC}19.2\% \\
    \textbf{None of the above} 
      & \cellcolor[HTML]{08306B}\textcolor{white}{100.0\%}
      & \cellcolor[HTML]{FFFFFF}0.0\%
      & \cellcolor[HTML]{FFFFFF}0.0\% \\
    \bottomrule
  \end{tabular}
\end{table}

\subsubsection{Support for SLT by Identify Framing}
We also examined how identity framings related to participants’ views on whether SLTs \textit{should} be developed (Table~\ref{tab:slt_support_identity}). Here, the contrast between cultural and disability framings becomes even clearer.

\begin{table}[h]
  \centering
  \caption{Participants' responses to ``Thinking about your answers about the possible benefits and harms of sign language technologies, should sign language technologies be developed?'' by their views on deafness.}
  \label{tab:slt_support_identity}
  \begin{tabular}{lccc}
    \toprule
     & \textbf{Yes} & \textbf{No} & \textbf{Other} \\
    \midrule
    \textbf{A cultural identity} 
      & \cellcolor[HTML]{1D5CD7}\textcolor{white}{76.9\%}
      & \cellcolor[HTML]{FFFFFF}0.0\%
      & \cellcolor[HTML]{A7C7FA}23.1\% \\
    \textbf{A disability} 
      & \cellcolor[HTML]{08306B}\textcolor{white}{100.0\%}
      & \cellcolor[HTML]{FFFFFF}0.0\%
      & \cellcolor[HTML]{FFFFFF}0.0\% \\
    \textbf{Both of the above} 
      & \cellcolor[HTML]{3E7EEB}61.5\%
      & \cellcolor[HTML]{E6EEFC}7.7\%
      & \cellcolor[HTML]{7CAEF7}30.8\% \\
    \textbf{None of the above} 
      & \cellcolor[HTML]{08306B}\textcolor{white}{100.0\%}
      & \cellcolor[HTML]{FFFFFF}0.0\%
      & \cellcolor[HTML]{FFFFFF}0.0\% \\
    \bottomrule
  \end{tabular}
\end{table}

Those who identified deafness as a \textit{disability} unanimously supported the development of SLTs (100\%). In contrast, participants who perceive deafness as a cultural identity exhibited more conditional support, with 76.9\% supporting the development. The most divided group were those identifying with \textit{both} perspectives, with 61.5\% supporting SLT development, 7.7\% opposing, and 30.8\% expressing conditional views.

These results illustrate how identity framings of deafness shape openness to SLTs. For participants grounded in cultural perspectives, SLTs may appear misaligned with Deaf cultural values. For those adopting disability-oriented or hybrid framings, SLTs are more readily interpreted as assistive technologies. This divergence suggests the importance of engaging with Deaf cultural frameworks when designing, deploying, or framing SLTs, rather than assuming uniform acceptance across communities.
\subsection{Policy Implications}
\label{sec:policy}
Policy contexts play a significant role in shaping how DHH communities experience and evaluate SLTs. Our survey explored both awareness of harmful and beneficial policies, as well as how exposure to sign language legislation influences perceptions of deaf identity.

\subsubsection{Policy Awareness}
\paragraph{Awareness of Harmful Policies.}
To better understand participants’ perceptions of systemic inequities, we asked whether they were aware of any policies or programs enacted in their region that have harmed deaf people. Responses were classified as \emph{Yes}, \emph{No}, or \emph{Unsure}. Table~\ref{tab:harmful_beneficial_policies_awareness} presents the proportion of participants within each audiological group who selected each response.

\begin{table}[h]
  \centering
  \caption{Awareness of Harmful vs. Beneficial Policies by Audiological Status (Percentages)}
  \label{tab:harmful_beneficial_policies_awareness}
  \begin{tabular}{lccc ccc}
    \toprule
    & \multicolumn{3}{c}{\textbf{Harmful}} & \multicolumn{3}{c}{\textbf{Beneficial}} \\
    \cmidrule(lr){2-4}\cmidrule(lr){5-7}
    & \textbf{Yes} & \textbf{No} & \textbf{Unsure} & \textbf{Yes} & \textbf{No} & \textbf{Unsure} \\
    \midrule
    \textbf{Deaf}
      & \cellcolor[HTML]{99BFF9}33.3\%
      & \cellcolor[HTML]{D9E6FB}19.0\%
      & \cellcolor[HTML]{2F72E9}47.6\%
      & \cellcolor[HTML]{6DA4F6}38.1\%
      & \cellcolor[HTML]{E6EEFC}14.3\%
      & \cellcolor[HTML]{2F72E9}47.6\% \\
    \textbf{HoH}
      & \cellcolor[HTML]{2F72E9}50.0\%
      & \cellcolor[HTML]{6DA4F6}33.3\%
      & \cellcolor[HTML]{E6EEFC}16.7\%
      & \cellcolor[HTML]{2F72E9}50.0\%
      & \cellcolor[HTML]{A7C7FA}25.0\%
      & \cellcolor[HTML]{A7C7FA}25.0\% \\
    \textbf{Hearing}
      & \cellcolor[HTML]{6DA4F6}33.3\%
      & \cellcolor[HTML]{A7C7FA}22.2\%
      & \cellcolor[HTML]{2F72E9}44.4\%
      & \cellcolor[HTML]{2F72E9}66.7\%
      & \cellcolor[HTML]{EAF1FE}11.1\%
      & \cellcolor[HTML]{A7C7FA}22.2\% \\
    \bottomrule
  \end{tabular}
\end{table}

Among deaf participants, 33.3\% reported awareness of harmful policies, while half (47.6\%) indicated uncertainty. In contrast, hard-of-hearing participants reported the highest rate of awareness, with half (50\%) selecting “Yes” and only 16.7\% selecting “Unsure”. These results suggest differences in how structural harms are recognized across audiological groups. Hard-of-hearing participants may occupy a position between deaf and hearing worlds, granting them exposure to systemic shortcomings. The relatively high rate of uncertainty among deaf individuals may reflect the underacknolwedgement of harmful policies due to the limited transparency or visibility. For hearing individuals, it suggests a lack of exposure to deaf communities.

\paragraph{Qualitative Accounts of Harmful Policies.}
To further contextualize the quantitative findings, participants who selected “Yes” to the question about harmful policies were invited to describe these policies in their own words. These open-ended responses provide deeper insight into the experiences on how systemic barriers are perceived across contexts. Their narratives revealed a range of structural and systemic barriers, which we organized into five prominent themes.

The most frequently cited concern involved \emph{inadequate access} to interpreting services, public services, or communication technologies (n=7). Participants described how access is often conditional, inconsistent, or reliant on informal practices rather than policy (P7, P16, P17). Participants pointed to failures in healthcare (P21), emergency systems (P10), and city planning (P7, P10, P12, P25):
\begin{quote}
    ``Use of VRI in hospitals reduced access for my friends. One friend went through cancer and was supposed to somehow look at the VRI screen while in extreme pain.’’ (P21)\\
    ``City events brought interpreters on stage without reserving a seating area for D/hh. Many venue seats were too far to see interpreters.’’ (P25)
\end{quote}

Several participants highlighted \emph{systemic failures in educational settings} (n=5), from the absence of trained educators (P10, P12, P30) to policies that inadvertently deprioritize Deaf culture (P12). One participant explained:
\begin{quote}
    \textit{``Children lack access in education and we are seeing a systemic issue of language deprivation... closure of Deaf schools, limited availability of educational interpreters.’’} (P12)
\end{quote}
Another noted the practical absence of ASL in public schools:
\begin{quote}
    \textit{``The Ontario government even added ASL classes to the high school curriculum, however with no one to teach, they have been ignored.’’} (P30)
\end{quote}

Respondents also discussed the \emph{lack of deaf representation in policy design} (n=4) and the top-down nature of decisions affecting the community:
\begin{quote}
    \textit{``Australia does not prioritize the voices of parents of deaf children or Deaf-led organisations... Policies are changing about us without us.’’} (P12)
\end{quote}
Other participants pointed to exclusions from disability benefits tied to the adoption of communication technologies grounded in a hearing-centric model, such as cochlear implants (P15, P23, P35).

A recurring tension emerged around the \emph{medicalization of deafness} (n=4). Several participants criticized policies that frame deafness primarily as a medical condition requiring intervention rather than recognizing it as a cultural identity (P15, P26, P35):
\begin{quote}
    \textit{``Deafness is still considered to be "hearing loss". For profoundly deaf people deafness is not their disability. Their real disability is blocks to accessing communication, language, education, society etc. Government etc still sees solutions to deafness to be amplified sounds.’’} (P26)
\end{quote}

Lastly, some responses referenced the \emph{historical and systematic structures} (n=3) that have marginalized deaf individuals:
\begin{quote}
    \textit{``Deaf people were also forcibly sterilized as recent as the late 1980s.’’} (P15)\\
    \textit{``Audism is rampant throughout policy... the limited definitions for accessing the Disability Tax Credit springs to mind.’’} (P23)
\end{quote}
These accounts suggest the deep historical roots of exclusion and the need to examine how current systems continue to perpetuate marginalization.

\paragraph{Awareness of Beneficial Policies.}
Awareness of beneficial policies was generally higher than harmful ones, though uneven across groups (Table~\ref{tab:harmful_beneficial_policies_awareness}). Two thirds of hearing (66.7\%) and half of hard-of-hearing (50\%) participants reported awareness of beneficial policies. In contrast, only 35\% of deaf participants indicated the same, a proportion roughly two-thirds that of the other groups. Moreover, 47.6\% of deaf participants selected “Unsure,” compared to just 22\% of hearing and 25\% of hard-of-hearing participants. These discrepancies suggest that deaf individuals, despite being the primary targets of such policies, may experience less visibility into their impact, or that those policies themselves are not perceived beneficial.

\paragraph{Qualitative Accounts of Beneficial Policies.}
Participants who reported awareness of beneficial policies were invited to describe them in open-ended responses. These responses revealed several recurring themes that offer insight into how DHH individuals perceive and experience policy level support.

The most frequently mentioned category was \emph{legislation frameworks} (n=9), including international, national, and regional laws designed to promote equity and access for disabled populations more broadly. One participant reflected positively on the expansion of access brought by these policies, while also noting the unintended consequences of educational placement decisions made under the Individuals with Disabilities Education Act (IDEA). Specifically, the policy's requirement to place students in the “least restrictive environment” was considered to be sometimes misaligned with the actual needs of deaf learners:
\begin{quote}
    \textit{``ADA was enacted during my lifetime and guaranteed access to public services. IDEA is a bit more controversial... worked out for me but sometimes put deaf students in inappropriately ‘least restrictive’ environments.’’} (P21)
\end{quote}

The second most prevalent theme was \emph{accessibility} (n=7). Participants described a range of practical improvements including televised sign language interpretation (P1), closed captioning mandates (P39), and video relay service programs (P15). While many highlighted that accessibility policies in theory are valuable, they often lack enforcement:
\begin{quote}
    \textit{``In theory, accessibility legislation should help, although it is not enforced so largely useless.’’} (P23)
\end{quote}
These reflections indicate an appreciation for the structural protections provided by law, while also acknowledging that they may be inappropriate or ineffective in practice.

Policies that supported \textit{workplace inclusion} also surfaced in the responses (n=2). One participant cited government mandates requiring a minimum quota of employees with disabilities and associated penalties for noncompliance (P9). Another highlighted the role of individual action in effecting institutional change:
\begin{quote}
    \textit{“The biggest was actually me standing up in my workplace which now pays for ASL classes. Walmart Canada funded my ASL 1–3 classes and ASL for first responders—which is more than our government has ever done.”} (P30)
\end{quote}

Finally, a few responses touched on \emph{financial supports and benefits} (n=2), such as discounted park and ski area admissions (P18), and support for accessible transportation (P35). These benefits contributed to feelings of inclusion in public life.

While participants valued these measures, they frequently emphasized the gap between formal policy and lived experience, with enforcement and implementation emerging as persistent concerns.

\subsubsection{Policy influence on Identity}
To explore whether national policy contexts influence perceptions related to our central theme of \emph{identity}, we categorized participants into two groups based on their country of residence history (as shown in Table~\ref{tab:freq}) and presence or absence of sign language legislation in those countries, as recognized by the World Federation of the Deaf (WFD) as of September 2024 \cite{wfd2020signlanguages}. Each participant’s country score was computed by assigning a fractional weight to each country they had lived in, proportionate to the number of countries reported. If a country had recognized sign language through formal legislation, it contributed positively to the participant's cumulative score. This approach allowed us classify the respondents to two comparative groups: those with a country score of 0 (indicating no exposure to sign language legislation; n=28), and those with a score greater than 0 (indicating residence in at least one country with such legislation; n=14).

Table~\ref{tab:deafness_perception_by_legislation} illustrates how participants’ perceptions of deafness vary depending on whether they have lived in countries with formal sign language legislation. Among those with no exposure to such policy environments (country score = 0), the majority (74.1\%) described deafness as both a cultural identity and a disability, with only 22.2\% identifying it solely as a cultural identity. In contrast, participants who had lived in at least one country with sign language legislation (country score $>$ 0) were more likely to view deafness exclusively as a cultural identity (50.0\%). Moreover, the only participant who described deafness exclusively as a disability belonged to the non-legislated group. These findings suggest that national policy contexts may shape how individuals understand and internalize deaf identity.

\begin{table}[h]
  \centering
  \caption{Perceptions of Deafness by Country Grouped by Sign Language Legislation}
  \label{tab:deafness_perception_by_legislation}
  \begin{tabular}{lcccc|c}
    \toprule
     & \textbf{Cultural} & \textbf{Disability} & \textbf{Both} & \textbf{None} & \textbf{n} \\
    \midrule
    \textbf{No legis.} 
      & \cellcolor[HTML]{A7C7FA}21.4\%
      & \cellcolor[HTML]{EAF1FE}3.6\%
      & \cellcolor[HTML]{2F72E9}75.0\%
      & \cellcolor[HTML]{FFFFFF}0.0\%
      & 28 \\
    \textbf{Legislation} 
      & \cellcolor[HTML]{6DA4F6}50.0\%
      & \cellcolor[HTML]{E6EEFC}7.1\%
      & \cellcolor[HTML]{99BFF9}35.7\%
      & \cellcolor[HTML]{E6EEFC}7.1\%
      & 14 \\
    \bottomrule
  \end{tabular}
\end{table}

\section{Discussion}
\subsection{Conditional Support and the Call for Deaf Leadership}
Our findings reveal that support for SLTs is contingent, not universal, and closely tied to broader patterns of linguistic access, power, and inclusion.
Support for SLTs is often conditional on whether DHH individuals are meaningfully involved in their development, with participants expressing concerns about hearing individuals dominating these spaces. While a majority of participants across audiological groups supported the idea of SLTs, DHH respondents were significantly more likely to select the “Other” option when asked whether such technologies should be developed, using the open-ended space to elaborate on their reservations. These responses reflected a deep concern that SLTs, when developed without deaf leadership, risk doing more harm than good. 

As evidenced throughout our analyses, hearing perspectives frequently diverged from those of DHH participants. Hearing respondents were more likely to uncritically endorse SLTs, often overemphasizing the utility of translation features, while DHH respondents highlighted concerns around cultural erasure, system inaccuracies, and the marginalization of DHH perspectives. These divergences suggest that \textit{hearing-dominated narratives} may obscure the needs of the communities SLTs intend to serve. 

Participants repeatedly invoked the principle of “\textit{nothing about us without us}”, to call attention to the ethical imperative that DHH individuals should lead the research, design, and deployment of SLTs. They warned that hearing-dominated SLT development, especially when devoid of linguistic or cultural fluency, often produces systems that are either ineffective or actively harmful. Several described such efforts as "vanity projects" or criticized them for siphoning resources away from initiatives that could more meaningfully benefit the DHH community. 



To develop technologies that are truly aligned with the needs and values of DHH individuals, researchers and technologists must center DHH perspectives from the outset. Only by centering DHH perspectives throughout the research and development pipeline can SLTs fulfill their promise as tools for accessibility, empowerment, and cultural affirmation. As our results have shown, these are requirements, rather than preferences, for many community members to support development of SLTs.

\subsection{Downstream Impacts of Sign Language Technology}
Participants also had hesitations about SLTs that may reflect existing structural and societal conditions. Rather than the technologies themselves, these concerns were centered around the downstream effects of deploying SLTs. In particular, participants were often concerned about access. Some were worried that SLTs would replace human interpreters. One participant discussed a situation where SLTs actually \emph{reduced} access, because VRI systems in healthcare required that a friend of theirs concentrate on a screen despite going through immense pain. Another participant described experiences where poor internet connection prevented them from using these technologies.

Other participants expressed concerns that SLTs would disincentivize hearing people from learning sign languages, with one participant worried that SLTs would prevent sign language learning from becoming normalized.

It is critical that SLTs are not seen as potential replacements for interpreters, particularly in situations where access is not possible without interpreters. Rather, SLTs should be supplementary in situations where interpreters cannot be accessed. Instituting safeguards that prevent this possibility are likely to reduce DHH communities' concerns about SLTs.

Further, it is important that all people are encouraged to learn sign language and provided with th tools to do so, particularly early in life. One participant pointed this out and described their experience with schools unable to staff a sign language program. Many other participants, especially hard-of-hearing participants, expressed regret that they had not learned sign language at an earlier age. 

\subsection{Implications for Inclusive and Ethical Design}
SLTs promise access, but when they are not developed ethically or with consideration for community needs, they can perpetuate marginalization
The findings of this study point to clear priorities for the inclusive and ethical design of SLTs. First, DHH individuals must be involved at every stage of the development process as researchers, designers, and decision-makers. Secondly, design efforts must also go beyond performance optimization. While system accuracy is critical, participants highlighted that performance alone does not determine the success or ethical alignment of a system. Cultural fluency and alignment with community values are essential dimensions of “good” system performance.

In addition, designers must consider the diversity of DHH experiences and signing practices. Not all DHH individuals have high sign language proficiency. Ensuring that SLTs are usable and accessible to individuals across this spectrum is essential to avoid reinforcing internal hierarchies or excluding already-marginalized subgroups within the DHH community. 

These findings call for a paradigm shift from building technologies for the DHH community to building them with the community. This not only improves system effectiveness, but also affirms the linguistic and cultural rights of DHH individuals and lays the foundation for technologies that can support equity, autonomy, and social inclusion. Centering deaf and hard of hearing leadership is not merely a best practice; rather, it is a precondition for equitable technology development.

\subsection{Limitations and Future Work}
One of the most significant challenges was recruitment. Reaching members of the DHH communities especially for those outside of dominant sign language communities (such as ASL communities) are frequently underrepresented in research due to systematic barriers in outreach and trust. Our final sample reflects the result of multiple iterations of recruitment efforts, including direct outreach to over 120 organizations and individuals across 35 countries, targeted posts in 39 Facebook groups for Deaf communities, engagement on platforms like Reddit, and the translation of the survey into ten spoken languages. Despite these efforts, our final sample size (n=42) reflects the difficulty of reaching a broad cross-section of the global DHH population. 

Nonetheless, we believe that the perspectives shared provide valuable insights into the diverse and often underutilized experiences of DHH individuals regarding SLTs. Rather than focusing solely on dominant sign languages or English-speaking contexts, we made a concerted effort to amplify the perspectives of DHH individuals from varied regions, backgrounds, and signing experiences.

Future studies could examine how perspectives shift over time as SLTs continue to evolve, or conduct in-depth qualitative interviews to capture experiences that surveys may not fully elicit. While limitations persist, we hope this study serves as a step toward more inclusive, deaf-centered research and design.
\section{Conclusion}
This study provides a global, multilingual analysis of DHH perspectives on SLTs. Insights from DHH individuals challenge the assumption that SLTs are inherently empowering by showing that support depends on structural inclusion and cultural
recognition.

We find that support for SLTs is deeply conditional, shaped by lived experience, linguistic access, and community inclusion. While many participants recognized the potential of SLTs to enhance accessibility and autonomy, concerns around cultural erasure, inaccurate performance, and the exclusion of deaf perspectives were also prominent, especially among those with higher sign language proficiency. Across our findings, DHH participants consistently emphasized that technologies must be developed with the communities they aim to serve, rather than designing them on behalf of those communities. DHH perspectives must be centered to ensure that SLTs are not only technically effective but also culturally affirming and empowering.

Further, participants frequently expressed concern that SLTs could \emph{reduce} access if they are used to replace interpreters or deployed in scenarios where they should not be. Participants also worried that SLTs could discourage people from learning sign languages. It is important that SLTs are implemented with caution, and that policies and safeguards are implemented to ensure that they are not used to replace interpreters or deployed inappropriately. It is also crucial that sign language learning is encouraged and facilitated for non-deaf people.

Centering DHH leadership is not only an ethical imperative, but is essential to ensuring that SLTs truly advance communication equity rather than reinforce existing hierarchies.

\bibliographystyle{ACM-Reference-Format}
\bibliography{sample-base}


\appendix
\section{Complete Survey}
\label{subsec:survey}
\textbf{Surveying Global Perspectives of Deaf and Hard-of-Hearing Individuals on Sign Language Technologies}
\subsection{Views of Sign Language Technology}
\label{subsec:views-slt}
\textbf{This survey uses the term, "sign language technology." What does this mean?} Technology related to sign language. Like sign language dictionaries, translation systems, video editing tools for signers, or educational tools that give students signing feedback.\\
- Which concerns do you have related to the development of sign language technologies? Please select all that apply (see above for a definition and examples of sign language technologies).\label{q1:concerns}\\
\hspace*{2em} - Hearing people making money from sign language technologies\\
\hspace*{2em} - System performance issues (for instance, worse performance for some users than others)\\
\hspace*{2em} - Few deaf people leading or contributing\\
\hspace*{2em} - System accessibility\\
\hspace*{2em} - Legal and political concerns (for instance, weakened legal protections for other accommodations)\\
\hspace*{2em} - No concerns\\
\hspace*{2em} - Other ()\\
- What benefits do you think could arise from the advancement of sign language technologies? Please check all that apply. \label{q2:benefits}\\
\hspace*{2em} - Greater availability of services (for instance, services that can be used in more locations or at the last minute)\\
\hspace*{2em} - More sign-language friendly technologies\\
\hspace*{2em} - Less need to read, write, or type spoken languages \\
\hspace*{2em} - Translations between primary sign language and other spoken or signed languages \\
\hspace*{2em} - The ability to use your own voice/signing avatar when being interpreted
\\
\hspace*{2em} - No benefits\\
\hspace*{2em} - Other\\
- Thinking about your answers about the possible benefits and harms of sign language technologies, should sign language technologies be developed?\\
\hspace*{2em} - Yes\\
\hspace*{2em} - No\\
\hspace*{2em} - Other\\
- Briefly describe the life experiences that have influenced your perspective on sign language technologies. For instance, how have sign language technologies themselves impacted you? Are there any situations, such as a lack of an interpreter, where sign language technologies may have been useful?\\
- Has your life experience shifted your opinion of sign language technologies in a positive, negative, or neutral way overall?\\
\hspace*{2em} - Positive\\
\hspace*{2em} - Neutral\\
\hspace*{2em} - Negative\\
- Are there any sign language technologies that have improved your overall educational experiences or increased your access to education as a student?\\
\hspace*{2em} - Yes\\
\hspace*{2em} - No\\
- If yes, describe or list these technologies.

\subsection{Views of Deafness}
- How do you view deafness?\\
\hspace*{2em} - A cultural identity\\
\hspace*{2em} - A disability\\
\hspace*{2em} - Both of the above\\
\hspace*{2em} - None of the above\\
- Are you aware of any policies or programs that have been enacted where you live that have harmed deaf people?\\
\hspace*{2em} - Yes\\
\hspace*{2em} - No\\
\hspace*{2em} - Unsure\\
- If yes, can you describe these policies or decisions?\\
- Are you aware of any policies or programs that have been enacted where you live that have benefited deaf people?\\
\hspace*{2em} - Yes\\
\hspace*{2em} - No\\
\hspace*{2em} - Unsure\\
- If yes, can you describe these policies or decisions?\\

\subsection{Demographics}
- Select all of the countries where you have lived\\
- Which best describes where you currently live?\\
\hspace*{2em} - Rural\\
\hspace*{2em} - Suburban\\
\hspace*{2em} - City\\
- What is your audiological status?\\
\hspace*{2em} - Deaf\\
\hspace*{2em} - Hard of hearing\\
\hspace*{2em} - Hearing\\
- Do you know a sign language?\\
\hspace*{2em} - Yes\\
\hspace*{2em} - No\\
- If yes, what is your primary sign language?\\
- Select your level of proficiency in your primary sign language, from 0 to 5 (adapted from the American Sign Language Proficiency Interview)\\
\hspace*{2em} - 0 - Unable to function in the language\\
\hspace*{2em} - 1 - Able to satisfy routine travel needs and minimum courtesy requirementsg\\
\hspace*{2em} - 2 - Able to satisfy routine social demands and limited work requirements\\
\hspace*{2em} - 3 - Able to sign with sufficient structural accuracy and vocabulary to participate effectively in most formal and informal conversations on practical, social, and professional\\
\hspace*{2em} - 4 - Able to use the language fluently and accurately on all level normally pertinent to professional needs\\
\hspace*{2em} - 5 - Language proficiency equivalent to that of a sophisticated native signer.\\
- What is your age?\\
- What is your gender?\\
\hspace*{2em} - Male\\
\hspace*{2em} - Female\\
\hspace*{2em} - Other\\

\section{Framing of deafness and identity}
To better understand how participants conceptualize deafness, we asked them to indicate whether they viewed deafness primarily as \emph{a cultural identity}, \emph{a disability}, \emph{both}, or \emph{none of the above}. This framing question aimed to explore how identity perspectives intersect with audiological status, sign language proficiency, and geographic context.

\subsubsection{Audiological Status.} 
Figure~\ref{tab:cdbn_audiological} displays the distribution of responses by audiological status. Among deaf participants, a majority (60\%) viewed deafness as both a cultural identity and a disability, while 35\% viewed it solely as a cultural identity. Only one deaf participant (5\%) selected “disability” alone. Hearing participants showed a more divided framing: 43\% selected “both,” another 43\% selected “cultural identity” alone, and 14\% selected “none of the above.” While the “both” option was the most common response across all groups, deaf participants were more likely than hearing or hard-of-hearing individuals to view deafness exclusively as a cultural identity. 

\begin{table}[h]
  \centering
  \caption{Perception of Deafness by Audiological Status (Percentages)}
  \label{tab:cdbn_audiological}
  \begin{tabular}{lcccc}
    \toprule
     & \textbf{Cultural Identity} & \textbf{Disability} & \textbf{Both} & \textbf{None} \\
    \midrule
    Deaf           & 35.0\% & 5.0\% & 60.0\% & 0.0\% \\
    Hard-of-hearing  & 17.0\% & 0.0\% & 83.0\% & 0.0\% \\
    Hearing        & 43.0\% & 0.0\% & 43.0\% & 14.0\% \\
    \bottomrule
  \end{tabular}
\end{table}

\subsubsection{Sign Language Proficiency.} Because prior literature suggests that sign language proficiency is closely linked to Deaf cultural identification \cite{ladd2003understanding}, we further analyzed identity framing across sign language proficiency groups (Table~\ref{tab:cdbn_asl}). This analysis revealed a clear trend: as sign language proficiency increased, so did the proportion of participants who identified deafness as a cultural identity (rising from 23\% in the low proficiency group to 43\% in the high proficiency group). Conversely, endorsement of the “both” framing declined from 75\% in the moderate proficiency group to 57\% in the high group. The sole participant who selected “disability” fell within the low proficiency group (level 2), further suggesting that sign language use may shape how participants conceptualize deafness, as a socialcultural experience rather than disability. This aligns with the claims made by \cite{10.1093/deafed/eni030}. Additionally, no participants with moderate or high SL proficiency viewed deafness as a disability alone or as “none of the above” which reinforces the strong association between language fluency and cultural affiliation.
\begin{table}[h]
  \centering
  \caption{Perception of Deafness by SL Proficiency Group (Percentages)}
  \label{tab:cdbn_asl}
  \begin{tabular}{lcccc}
    \toprule
    \textbf{SL Proficiency} & \textbf{Cultural Identity} & \textbf{Disability} & \textbf{Both} & \textbf{None} \\
    \midrule
    Low (0--2) & 23.0\% & 8.0\% & 62.0\% & 8.0\% \\
    Moderate (3--4) & 25.0\% & 0.0\% & 75.0\% & 0.0\% \\
    High (5)    & 43.0\% & 0.0\% & 57.0\% & 0.0\% \\
    \bottomrule
  \end{tabular}
\end{table}

\subsubsection{Residential Context.}
Finally, we examined whether participants’ area of residence, categorized as rural, suburban, or city, was associated with their framing of deafness. We hypothesized that urban participants, who may have greater access to deaf communities and resources, would be more likely to identify deafness as a cultural identity. However, the distribution of identity framings was relatively consistent across geographic contexts. Cultural identity alone was endorsed by 38\% of rural participants, 29\% of suburban participants, and 29\% of city participants. These results suggest that geographic location may play a limited role in shaping how individuals frame deafness, at least within the context of this sample.
\begin{table}[h]
  \centering
  \caption{Perception of Deafness by Area of Residence (Percentages)}
  \label{tab:cdbn_residence}
  \begin{tabular}{lcccc}
    \toprule
    \textbf{Residential} & \textbf{Cultural Identity} & \textbf{Disability} & \textbf{Both} & \textbf{None} \\
    \midrule
    Rural     & 38.0\% & 0.0\% & 50.0\% & 13.0\% \\
    Suburban  & 29.0\% & 0.0\% & 71.0\% & 0.0\% \\
    City      & 29.0\% & 6.0\% & 65.0\% & 0.0\% \\
    \bottomrule
  \end{tabular}
\end{table}

These findings suggest the influence of sign language use and audiological identity on how individuals conceptualize deafness. While the “both” framing remains prominent, higher sign language proficiency is associated with stronger cultural identification. In contrast, residential context appeared to exert minimal influence, which suggests that identity framing may be more associated with linguistic and social experience than with geographic environment.

\section{Opinion of SLT by Audiological Status}
Participants were also asked: “Has your life experience shifted your opinion of sign language technologies in a positive, negative, or neutral way overall?” As shown in Table~\ref{tab:life-shift}, responses were mixed across groups. Deaf respondents were evenly split across positive (25\%), negative (25\%), and neutral (50\%) views. Hard-of-hearing participants were more likely to report a positive shift (66.7\%), while Hearing participants were divided evenly between positive and neutral (42.9\% each), with a smaller portion reporting negative experiences (14.3\%).
\begin{table}[h]
  \centering
  \caption{Responses to: ``Has your life experience shifted your opinion of sign language technologies in a positive, negative, or neutral way overall?''}
  \begin{tabular}{lccc}
    \toprule
     & \textbf{Positive} & \textbf{Neutral} & \textbf{Negative} \\
    \midrule
    Deaf & 25.0\% & 50.0\% & 25.0\% \\
    Hard-of-hearing & 66.7\% & 16.7\% & 16.7\% \\
    Hearing & 42.9\% & 42.9\% & 14.3\% \\
    \bottomrule
  \end{tabular}
  \label{tab:life-shift}
\end{table}
\end{document}